\documentclass[prd,amsfonts]{revtex4}
\usepackage{graphicx}% Include figure files
\usepackage{bm}% bold math
\begin{document}

\title{Oppenheimer-Volkoff Equation in Relativistic MOND Theory}

\author{Xing-hua Jin}
\author{Xin-zhou Li}
\email{kychz@shtu.edu.cn(X.Z.Li)}

\affiliation{Shanghai United Center for Astrophysics (SUCA),
Shanghai Normal University, 100 Guilin Road, Shanghai 200234, China}

\date{\today}
% It is always \today, today,
%  but any date may be explicitly s
\begin{abstract}

In this paper, we discuss the internal and external metric of the
semi-realistic stars in relativistic MOND theory. We show the
Oppenheimer-Volkoff equation in relativistic MOND theory and get the
metric and pressure inside the stars to order of post-Newtonian
corrections. We study the features of motion around the static,
spherically symmetric stars by Hamilton-Jacobi mothod, and find
there are only some small corrections in relativistic MOND theory.

\end{abstract}

\maketitle


\vspace{0.4cm} \noindent\textbf{1. Introduction} \vspace{0.4cm}

The "missing mass" problem comes of the discrepancy between two
methods which relate to the measurements of mass to luminosity
ratio. One is to measure the cluster's mass to luminosity by dynamic
method; and the other is to directly measure the mass to luminosity
through the rotation curve of spiral galaxies. To solve the problem,
a prevalent way is to assume that there is much more invisible
non-baryonic dark matter than visible baryonic matter in the
universe. However, as an alternative to dark matter,
Milgrom\cite{Milgrom} proposed a Modified Newtonian Dynamics(MOND)
which implies that Poisson equation should be rewritten as
$\nabla\left(\nabla\Phi{|\nabla\Phi|}/{a_0}\right)=4\pi G\rho$ when
$|\nabla\Phi|\ll a_0$, where $a_0\approx 1\times 10^{-8}cm s^{-2}$
from empirical data. One found that MOND can explain not only the
rotation curve\cite{Sarders} of spiral galaxies but also the
Tully-Fisher law\cite{Tully}: $L_K\propto v_a^4$, where $L_K$ is the
infrared luminosity of a disk galaxy and $v_a$ is the asymptotic
rotational velocity.

Though MOND can explain the experiment successfully, it is not a
theory and there exists some theoretical problems, such as the
conservation of momentum and angular momentum. Recently,
Bekenstein\cite{B1} has developed MOND to be a relativistic theory
of gravity, abbreviated TeVeS, which depends on a tensor field
$g_{\alpha\beta}$, a vector field $\mathfrak{U}_{\alpha}$, a
dynamical scalar field $\phi$ and non-dynamical scalar field
$\sigma$. The theory also involves a free function $F$, a length
scale $\ell$ and two positive dimensionless parameters, $k$ and $K$.
Bekenstein has shown that TeVeS has general relativity(GR) as its
limit when $k\rightarrow 0$ with $\ell\propto k^{-3/2}$ and
$K\propto k$ in Friedmann-Robertson-Walker(FRW) cosmology, and TeVeS
has a MOND and Newtonian limit under the proper circumstance. TeVeS
also passes the elementary solar system tests of gravity theory
\cite{B2}. Even though TeVeS is not a perfect theory, such as the
violation of the local Lorentz invariance and the free choosing of
the function $F$, TeVeS is still strong interesting recently. A
recent paper \cite{Skordis} discussed the large scale struture in
TeVeS theory which show that it may be possible to reproduce
observations of the cosmic microwave background and galaxy
distributions.

The external spacetime of the spherically symmetric objects in TeVeS
has been explored by Bekenstein\cite{B1} and
Giannios\cite{Giannios}. Bekenstein has studied the post-Newtonian
corrections of the static, spherically symmetric metric outside the
spherically symmetric objects, and Giannios has derived the analytic
expression for the physical metric. Up to now, there is no research
on the stellar structure in TeVeS. In this paper we study
analytically a semi-realistic star to get the primary characteristic
of stellar structure in TeVeS. Firstly, we discuss the asymptotic
vacuum spacetimes in the exterior of the stars by use of a static,
spherically symmetric metric which can be transformed into the
metric Bekenstein adopted. Secondly, we consider the
Oppenheimer-Volkoff equation in TeVeS, and use the series expansion
to compute the metric and pressure inside the stars. Finally, to get
the observable correct effect of TeVeS, we explore the features of
motion around the spherically symmetric stars by Hamilton-Jacobi
mothod.

\vspace{0.4cm} \noindent\textbf{2. The Brief Review of TeVeS}
\vspace{0.4cm}

The basic equations of TeVeS can be derived from the
action\cite{B1}:
\begin{equation}
S=S_g+S_v+S_s+S_m,
\end{equation}
where the actions of the tensor field($S_g$), the vector
field($S_v$), the scalar field($S_s$) and the matter($S_m$) are:
\begin{equation}
S_g=(16\pi G)^{-1}\int g^{\alpha\beta} R_{\alpha\beta} (-g)^{1/2}
d^4 x, \label{EH}
\end{equation}

\begin{equation}
S_s =-{\scriptstyle 1\over\scriptstyle 2}\int\big[\sigma^2
h^{\alpha\beta}\phi_{,\alpha}\phi_{,\beta}+{\scriptstyle
1\over\scriptstyle 2}G \ell^{-2}\sigma^4 F(kG\sigma^2)
\big](-g)^{1/2} d^4 x, \label{scalar}
\end{equation}

\begin{equation}
S_v =-{K\over 32\pi G}\int \big[g^{\alpha\beta}g^{\mu\nu}
\mathfrak{U}_{[\alpha,\mu]} \mathfrak{U}_{[\beta,\nu]}
-2(\tilde{\lambda}/K)(g^{\mu\nu}\mathfrak{U}_\mu \mathfrak{U}_\nu
+1)\big](-g)^{1/2} d^4 x, \label{vector}
\end{equation}

\begin{equation}
S_m=\int \mathcal{L}(\tilde g_{\mu\nu}, f^\alpha,
f^\alpha{}_{|\mu}, \, \cdots)(-\tilde g)^{1/2} d^4x. \label{matter}
\end{equation}
In the equations above, $g$ is the determinant of Einstein metric
$g_{\alpha\beta}$, $R_{\alpha\beta}$ is the Ricci tensor of
$g_{\alpha\beta}$ just as in GR,
$h^{\alpha\beta}=g^{\alpha\beta}-\mathfrak{U}^{\alpha}\mathfrak{U}^{\beta}$,
and $\tilde{g}_{\alpha\beta}$ is the physical metric which is
related to the Einstein metric by $\tilde g_{\alpha\beta} =
e^{-2\phi} (g_{\alpha\beta}+\mathfrak{U}_\alpha \mathfrak{U}_\beta)
- e^{2\phi}\mathfrak{U}_\alpha \mathfrak{U}_\beta$. $F(\mu)$ is a
free dimensionless function, with $\mu=kG\sigma^2$.
$\tilde{\lambda}$ is a spacetime dependent Lagrange multiplier and
$f^{\alpha}$ symbolically represents the matter fields. The symbol
$|$ denotes the covariant derivatives with respect to
$\tilde{g}_{\mu\nu}$ and a pair of indices surrounded by brackets
stands for antisymmetrization, i.e.
$A_{[\mu}B_{\nu]}=A_{\mu}B_{\nu}-A_{\nu}B_{\mu}$.

Varying the action $S$ with respect to $g^{\alpha\beta}$,
$\mathfrak{U}_\alpha$ and $\phi$, one will get the metric equation
\begin{equation}\label{gab}
G_{\alpha\beta} = 8\pi G\Big[\tilde T_{\alpha\beta} +(1-e^{-4\phi})
\mathfrak{U}^\mu \tilde T_{\mu(\alpha} \mathfrak{U}_{\beta)}
+\tau_{\alpha\beta}\Big]+ \Theta_{\alpha\beta},
\label{gravitationeq}
\end{equation}
the vector equation
\begin{equation}
K\mathfrak{U}^{[\alpha;\beta]}{}_{;\beta}+\tilde{\lambda}
\mathfrak{U}^\alpha+8\pi G\sigma^2
\mathfrak{U}^\beta\phi_{,\beta}g^{\alpha\gamma}\phi_{,\gamma} = 8\pi
G (1-e^{-4\phi}) g^{\alpha\mu} \mathfrak{U}^\beta \tilde
T_{\mu\beta}, \label{vectoreq}
\end{equation}
and the scalar equation
\begin{equation}
\left[\mu h^{\alpha\beta}\phi_{,\alpha} \right]_{;\beta}=
kG\big[g^{\alpha\beta} + (1+e^{-4\phi}) \mathfrak{U}^\alpha
\mathfrak{U}^\beta\big] \tilde T_{\alpha\beta}, \label{s_equation}
\end{equation}
where
\begin{equation}
\tau_{\alpha\beta}\equiv
\sigma^2\Big[\phi_{,\alpha}\phi_{,\beta}-{\scriptstyle 1\over
\scriptstyle 2}g^{\mu\nu}\phi_{,\mu}\phi_{,\nu}\,g_{\alpha\beta}-
\mathfrak{U}^\mu\phi_{,\mu}\big(\mathfrak{U}_{(\alpha}\phi_{,\beta)}-
{\scriptstyle 1\over \scriptstyle
2}\mathfrak{U}^\nu\phi_{,\nu}\,g_{\alpha\beta}\big)\Big] \label{tau}
-{\scriptstyle 1\over\scriptstyle 4}G \ell^{-2}\sigma^4 F(\mu)
g_{\alpha\beta},
\end{equation}
\begin{equation}
\Theta_{\alpha\beta}\equiv
K\Big(g^{\mu\nu}\mathfrak{U}_{[\mu,\alpha]}
\mathfrak{U}_{[\nu,\beta]} -{\scriptstyle 1\over \scriptstyle 4}
g^{\sigma\tau}g^{\mu\nu}\mathfrak{U}_{[\sigma,\mu]}
\mathfrak{U}_{[\tau,\nu]}\,g_{\alpha\beta}\Big)- \tilde{\lambda}
\mathfrak{U}_\alpha\mathfrak{U}_\beta, \label{Theta}
\end{equation}
and a pair of indices surrounded by parenthesis stands for
symmetrization, i.e.
$A_{(\mu}B_{\nu)}=A_{\mu}B_{\nu}+A_{\nu}B_{\mu}$. If one models the
matter as a perfect fluid, then the energy-momentum tensor has the
form
\begin{equation}
\tilde T_{\alpha\beta}=\tilde\rho \tilde u_\alpha\tilde u_\beta
+\tilde p(\tilde g_{\alpha\beta}+\tilde u_\alpha \tilde u_\beta)
\label{oldT},
\end{equation}
where $\tilde\rho$ is the proper energy density, $\tilde p$ the
pressure and $\tilde u_\alpha$ the 4-velocity, with $\tilde
u_\alpha=e^{\phi}\,\mathfrak{U}_\alpha$, all three expressed in the
physical metric.
 If varying the action $S_s$ with respect to $\sigma$, one can
arrive at the equation, $-\mu F(\mu ) -{\scriptscriptstyle 1\over
\scriptscriptstyle 2}\, \mu ^2F'(\mu ) = k\ell^2
h^{\alpha\beta}\phi_{,\alpha}\phi_{,\beta}$. There is some
discussion about the free function $F(\mu)$ in \cite{B1,Giannios}.
Because in this paper, we are interested in the post-Newtonian
corrections not the purely MOND case, we shall take $\mu=1$, i.e.
$\sigma^2=1/(kG)$.

\vspace{0.4cm} \noindent\textbf{3. External Spacetime of a Static,
Spherically Symmetric Star in TeVeS} \vspace{0.4cm}

The metric of the static and spherically symmetric stars can be
written as
\begin{equation}\label{ds1}
ds^2=g_{\alpha\beta}dx^{\alpha}dx^{\beta}
=-e^{\nu(r)}dt^2+e^{\lambda(r)}dr^2+r^2(d\theta^2+\sin^2\theta
d\varphi^2),
\end{equation}
where $\nu$ and $\lambda$ are the function of $r$ only. It is worth
noting that metric (\ref{ds1}) has the different form with that in
Ref.\cite{B1}, but both metric can be transformed between each
other. As we are looking for the static solutions, we will take the
vector field to be pointing in the timelike direction. Normalized to
$\mathfrak{U}_\alpha\mathfrak{U}^\alpha=-1$, it has
\begin{equation}\label{un}
\mathfrak{U}^\alpha=\{e^{-\nu/2},0,0,0\}. \label{U*}
\end{equation}
Using Eqs.(\ref{gab}) and (\ref{un}), we get easily the components
of gravitational equation. The $tt$ component is:
\begin{equation}\label{tt}
\frac{e^{-\lambda}}{r^2}\big(-1+e^{\lambda}+r\lambda'\big)=8\pi
G\big[\tilde{\rho}e^{2\phi}+\frac{1}{kG}\frac{e^{-\lambda}}{2}(\phi')^2\big]
+Ke^{-\lambda}\big[\frac{(\nu')^2}{8}+\frac{\nu''}{2}-\frac{\nu'\lambda'}{4}+\frac{\nu'}{r}\big],
\end{equation}
the $rr$ component is:
\begin{equation}\label{rr}
\frac{e^{-\lambda}}{r^2}\big(1-e^{\lambda}+r\nu'\big)=8\pi
G\big[\tilde{p}e^{-2\phi}+\frac{1}{kG}\frac{e^{-\lambda}}{2}(\phi')^2\big]
+Ke^{-\lambda}\big[-\frac{(\nu')^2}{8}\big],
\end{equation}
and the $\theta\theta$ component is:
\begin{equation}\label{thth}
\frac{e^{-\lambda}}{2}\big[\frac{(\nu')^2}{2}+\nu''-\frac{\nu'\lambda'}{2}+\frac{\nu'-\lambda'}{r}\big]=8\pi
G\big[\tilde{p}e^{-2\phi}-\frac{1}{kG}\frac{e^{-\lambda}}{2}(\phi')^2\big]
+Ke^{-\lambda}\big[\frac{(\nu')^2}{8}\big].
\end{equation}
The $\varphi\varphi$ component is proportional to the $\theta\theta$
component, so there is no need to consider it separately. In
principle, the first two components of gravitational equations are
enough to solve the metric. Furthermore the scalar equation becomes
\begin{equation}\label{phip}
\big[\frac{\phi'r^2}{e^{(\lambda-\nu)/2}}\big]'=kG(\tilde{\rho}+3\tilde{p})e^{(\nu+\lambda)/2}e^{-2\phi}r^2.
\end{equation}
Similar to Bekenstein's definition\cite{B1}, we define the "scalar
mass" $m_s$ as
\begin{equation}\label{ms}
m_s=m_s(R)=4\pi\int_0^R(\tilde{\rho}+3\tilde{p})e^{(\nu+\lambda)/2}e^{-2\phi}r^2dr,
\end{equation}
where $R$ is the radius of a spherical symmetric star.

We consider the case that the exterior of the star is vacuum, and
have $\tilde{\rho}=0$ and $\tilde{p}=0$. For $r>R$, Eq.(\ref{phip})
becomes
\begin{equation}
\phi'=k\frac{Gm_s}{4\pi}\frac{e^{(\lambda-\nu)/2}}{r^2}.
\end{equation}
Substitute $\tilde{\rho}$, $\tilde{p}$ and $\phi'$ into
Eqs.(\ref{tt}) and (\ref{rr}), we have
\begin{equation}\label{tt1}
\frac{e^{-\lambda}}{r^2}\big(-1+e^{\lambda}+r\lambda'\big)=
k\frac{(Gm_s)^2}{4\pi}\frac{e^{-\nu}}{r^4}
+Ke^{-\lambda}\big[\frac{(\nu')^2}{8}+\frac{\nu''}{2}-\frac{\nu'\lambda'}{4}+\frac{\nu'}{r}\big],
\end{equation}
and
\begin{equation}\label{rr1}
\frac{e^{-\lambda}}{r^2}\big(1-e^{\lambda}+r\nu'\big)=
k\frac{(Gm_s)^2}{4\pi}\frac{e^{-\nu}}{r^4}
+Ke^{-\lambda}\big[-\frac{(\nu')^2}{8}\big].
\end{equation}
For getting the post-Newtonian corrections, one can expand $e^{\nu}$
and $e^{\lambda}$ as
\begin{equation}\label{nu}
e^{\nu}=\alpha_0\big[1-\frac{r_e}{r}+\alpha_2\big(\frac{r_e}{r}\big)^2+\alpha_3\big(\frac{r_e}{r}\big)^3+\ldots\big],
\end{equation}
\begin{equation}
e^{\lambda}=\beta_0\big[1+\beta_1\big(\frac{r_e}{r}\big)+\beta_2\big(\frac{r_e}{r}\big)^2+\beta_3\big(\frac{r_e}{r}\big)^3+\ldots\big],
\end{equation}
where $r_e$ is a length scale to be determined in the next section,
and the size of the coefficient of the $r_e/r$ term in Eq.(\ref{nu})
has been absorbed into $r_e$. Substituting them into Eqs.(\ref{tt1})
and (\ref{rr1}), matching the coefficient of like powers of $1/r$,
and solving the recurrence relation, we obtain
\begin{eqnarray}
\alpha_2&=&0,
\\\nonumber
\alpha_3&=&-\frac{K}{48}+\frac{k}{a_0}\frac{(Gm_s)^2}{4\pi}\frac{1}{6r_e^2},
~\ldots
\end{eqnarray}

\begin{eqnarray}\label{be}
\beta_0&=&1,
\\\nonumber
\beta_1&=&1,
\\\nonumber
\beta_2&=&1+\frac{K}{8}-\frac{k}{a_0}\frac{(Gm_s)^2}{4\pi}\frac{1}{r_e^2},
\\\nonumber
\beta_3&=&1+\frac{5K}{16}-\frac{k}{a_0}\frac{(Gm_s)^2}{4\pi}\frac{5}{2r_e^2},~\ldots
\end{eqnarray}
To gain the value of the coefficient of $\alpha_0$, we can use the
fact that the limit of TeVeS is GR when $k\rightarrow 0$ and
$K\rightarrow 0$. Comparing the expansion with the Schwarzschild
metric in GR, we get $\alpha_0=1$. After substituting the expansion
into Eq.(\ref{phip}) and integrating the equation, we have
\begin{equation}\label{phie}
\phi=\phi_c-\frac{kGm_s}{4\pi}\frac{1}{r}-\frac{kGm_sr_e}{8\pi}\frac{1}{r^2}+\mathcal{O}(r^{-3}),
\end{equation}
where the constant $\phi_c$ is a cosmological value and can be
absorbed in the rescaling $t$ and $r$ coordinates.

Because of the relation $\tilde g_{tt}=-e^{2\phi+\nu}$, $\tilde
g_{rr}=e^{-2\phi+\lambda}$ and $\tilde g_{\theta\theta}=\tilde
g_{\varphi\varphi}/\sin^2{\theta}=r^2e^{-2\phi}$, the physical
metric becomes
\begin{eqnarray}
\tilde{g}_{tt}&=&-1+\frac{2G_Nm}{r}-\frac{2G_Nm(G_Nm-\frac{r_e}{2})}{r^2}+\mathcal{O}(r^{-3}),
\\\nonumber
\tilde{g}_{rr}&=&1+\frac{2G_Nm}{r}+\frac{2G_Nm(G_Nm+\frac{r_e}{2})-k\frac{(Gm_s)^2}{4\pi}+\frac{K}{8}r_e^2}{r^2}+\mathcal{O}(r^{-3}),
\\\nonumber
\tilde{g}_{\theta\theta}&=&r^2+(2G_Nm-r_e)r+2G_Nm(G_Nm-\frac{r_e}{2})+\mathcal{O}(r^{-1}),
\\\nonumber
\tilde{g}_{\varphi\varphi}&=&\sin^2\theta\tilde{g}_{\theta\theta},
\end{eqnarray}
where $2G_Nm=r_e+2k\frac{Gm_s}{4\pi}$.

\vspace{0.4cm} \noindent\textbf{4. Oppenheimer-Volkoff equation in
TeVeS} \vspace{0.4cm}

To get some characteristic of stellar structure in TeVeS, we will
research a semi-realistic stars in this section. In other words we
can get the kernel information about the stellar structure in TeVeS
through the simplified model, and to find how difference the stellar
structure is between in TeVeS and in GR. The simplified model of
star comes from assuming that the fluid is incompressible: the
density is a constant $\tilde{\rho}_0$ at $r<R$ and it vanishes at
$r>R$. In the internal of a star with ideal fluid
$\tilde{T}_{\alpha\beta}$, the $r$ component of energy-momentum
conservation law $\tilde{T}^{\alpha\beta}_{~~;\beta}=0$ gives
\begin{equation}\label{tab}
\frac{d\big(\tilde{p}e^{-2\phi}\big)}{dr}=-\frac{\big(\tilde{p}e^{-2\phi}+\rho
e^{2\phi}\big)}{2}\frac{d\nu}{dr}.
\end{equation}

Rewriting Eq.(\ref{tt}) and Eq.(\ref{rr}), we have
\begin{equation}\label{ttr}
\frac{e^{-\lambda}}{r^2}\big(-1+e^{\lambda}+r\lambda'\big)=8\pi
G\bar{\rho},
\end{equation}
and
\begin{equation}\label{rrr}
\frac{e^{-\lambda}}{r^2}\big(1-e^{\lambda}+r\nu'\big)=8\pi G\bar p.
\end{equation}
where
$\bar\rho=\tilde{\rho}e^{2\phi}+\frac{1}{kG}\frac{e^{-\lambda}}{2}(\phi')^2
+\frac{K}{8\pi
G}e^{-\lambda}\big[\frac{(\nu')^2}{8}+\frac{\nu''}{2}-\frac{\nu'\lambda'}{4}+\frac{\nu'}{r}\big]$
and $\bar
p=\big[\tilde{p}e^{-2\phi}+\frac{1}{kG}\frac{e^{-\lambda}}{2}(\phi')^2\big]
+\frac{K}{8\pi G}e^{-\lambda}\big[-\frac{(\nu')^2}{8}\big]$.
According to Eq.(\ref{ttr}), it is convient to replace $\lambda(r)$
with a new function $m(r)$, given by
\begin{equation}
m(r)=\frac{1}{2G}\left(r-re^{-\lambda}\right),
\end{equation}
or equivalently
\begin{equation}
e^{-\lambda}=1-\frac{2Gm(r)}{r}.
\end{equation}
Then Eq.(\ref{ttr}) becomes
\begin{equation}
\frac{dm(r)}{dr}=4\pi r^2\bar\rho,
\end{equation}
which can be integrated to obtain
\begin{equation}
m(r)=4\pi\int_0^r\bar\rho(r')r'^2dr'.
\end{equation}
For $r\geq R$,
\begin{equation}\label{lam}
e^{-\lambda}=1-\frac{2
G\left[4\pi\int_0^R\bar\rho(r')r'^2dr'+4\pi\int_R^r\bar\rho(r')r'^2dr'\right]}{r}=1-\frac{2Gm_g(R)}{r}
-\frac{8\pi G\int_R^r\bar\rho(r')r'^2dr'}{r},
\end{equation}
where $m_g$ is the "gravitational mass", with
$m_g=m_g(R)=4\pi\int_0^R\bar\rho(r')r'^2dr'$, and in the integration
the expression of $\bar\rho(r')$ depends on the internal metric
which has a complex form. At $r=R$, using Eqs.(\ref{rr}), (\ref{be})
and (\ref{lam}), we have the relation between $m_s$, $m_g$ and $r_e$
\begin{equation}
r_e-{K r_e{}^2\over 8R}+{kG^2m_s{}^2\over 4\pi
R}+\mathcal{O}\left(\frac{r_e{}^3}{R^2}\right)=2Gm_g.
\end{equation}
In terms of $m(r)$, Eq.(\ref{rrr}) can be rewritten
\begin{equation}\label{dnu}
\frac{d\nu}{dr}=\frac{2Gm(r)+8\pi G\bar p
r^3}{r\left[r-2Gm(r)\right]}.
\end{equation}
Combining Eq.(\ref{dnu}) with Eq.(\ref{tab}) allows us to obtain the
Oppenheimer-Volkoff equation\cite{Oppen} in TeVeS
\begin{equation}
\frac{d\big(\tilde{p}e^{-2\phi}\big)}{dr}=-(\tilde{p}e^{-2\phi}+\tilde\rho
e^{2\phi})\frac{Gm(r)+4\pi G\bar pr^3}{r\left[r-2Gm(r)\right]}.
\end{equation}

To post-Newtonian corrections, in the interiors of static, spherical
symmetry stars, we expand $e^{\nu}$, $e^{\lambda}$, $\phi$ and
$\tilde p$ as
\begin{equation}\label{nur}
e^{\nu}=a_0\big[1+\big(\frac{r}{r_i}\big)^2+a_2\big(\frac{r}{r_i}\big)^4+a_3\big(\frac{r}{r_i}\big)^6+\ldots\big],
\end{equation}
\begin{equation}
e^{\lambda}=b_0\big[1+b_1\big(\frac{r}{r_i}\big)^2+b_2\big(\frac{r}{r_i}\big)^4+b_3\big(\frac{r}{r_i}\big)^6+\ldots\big],
\end{equation}
\begin{equation}
\tilde{p}=c_0\big[1+c_1\big(\frac{r}{r_i}\big)^2+c_2\big(\frac{r}{r_i}\big)^4+c_3\big(\frac{r}{r_i}\big)^6+\ldots\big],
\end{equation}
\begin{equation}\label{phii}
\phi=d_1\big(\frac{r}{r_i}\big)^2+d_2\big(\frac{r}{r_i}\big)^4+d_3\big(\frac{r}{r_i}\big)^6+\ldots,
\end{equation}
where $r_i$ is a length scale to be determined later and the size of
the coefficient of the $r/r_i$ term in Eq.(\ref{nur}) has been
absorbed into $r_i$. In Eq.(\ref{phii}), we need not write down the
constant term because the constant is a cosmological value and can
be absorbed in the rescaling $t$ and $r$ coordinates. Substituting
them into Eq.(\ref{tt}), Eq.(\ref{rr}), Eq.(\ref{phip}) and
Eq.(\ref{tab}), matching the coefficient of like powers of $r^2$,
and solving the recurrence relation, we have
\begin{equation}
a_2 = \frac{27k(2 - K)}{80\pi} + \frac{8\pi G\tilde{\rho}_0(5 -
4K)r_i^2}{30(2 - K)} +
    \frac{5 - 4K^2 + 6kG\tilde{\rho}_0r_i^2 + K(8 - 3kG\tilde{\rho}_0r_i^2)}{10(2 - K)},~\ldots
\end{equation}

\begin{eqnarray}\label{bs}
b_0&=&1,
\\\nonumber
b_1&=&K+\frac{8\pi G\tilde{\rho_0}}{3}r_i^2,
\\\nonumber
b_2 &=& \frac{K^3(16 \pi - 9 k)}{16\pi (-2 + K)} + \frac{K^2(-348\pi
+ 243k + 640\pi^2G r_i^2 \tilde{\rho}_0)}{120\pi (-2+K)}
\\\nonumber
    &-&\frac{K [-243k - 1632\pi^2Gr_i^2\tilde{\rho}_0 +
              1280\pi^3G^2 r_i^4\tilde{\rho}_0^2 +72\pi(2 + 3kG r_i^2\tilde{\rho}_0)]}{180\pi (-2 + K)} -
    \frac{1280\pi^3G^2 \tilde{\rho}_0^2 +
          27k(3 + 8\pi Gr_i^2\tilde{\rho}_0)}{90\pi (-2 + K)},~\ldots
\end{eqnarray}

\begin{eqnarray}
c_0&=&\frac{\tilde{\rho_0}}{3}+\frac{2-K}{8 \pi Gr_i^2},
\\\nonumber
c_1&=&\frac{1}{3}+\frac{kG(2-K)}{16\pi G} +\frac{(10-5K)[8\pi
G-3kG(2-K)]}{48\pi G(-2+K+8\pi G\tilde{\rho_0}r_i^2)},
\\\nonumber
c_2 &=& \frac{27k^2(-2 + K)^3}{128\pi^2(-6 + 3 K +
              8 \pi G r_i^2\tilde{\rho}_0)} +
    \frac{32 (-5 + 4 K)\pi^2G^2r_i^4\tilde{\rho}_0^2}{15(-2 + K)(-6 + 3 K +
              8 \pi G r_i^2\tilde{\rho}_0)}
\\\nonumber
              &-&\frac{4\pi G r_i^2\tilde{\rho}_0[-15 + 48k\pi G r_i^2\tilde{\rho}_0 +
              K (27 - 44kGr_i^2\tilde{\rho}_0) +
              2 K^2 (-3 + 5kGr_i^2\tilde{\rho}_0)]}{15(-2 + K)(-6 + 3 K +
              8 \pi G r_i^2\tilde{\rho}_0)}
              \\\nonumber
              &-&
    \frac{3k(-2 + K)^2(5 K - 2 (9 + 5 kG r_i^2\tilde{\rho}_0))}{40\pi(-6 +
              3 K + 8 \pi G r_i^2\tilde{\rho}_0)}
              \\\nonumber
              &-&
    \frac{60 - 492kG r_i^2\tilde{\rho}_0 + 40k^2 G^2r_i^4\tilde{\rho}_0^2 +
          8K^2(3 + 10 kG r_i^2\tilde{\rho}_0) +
          K (-93 + 86kG r_i^2\tilde{\rho}_0 -
                20k^2 G^2 r_i^4\tilde{\rho}_0^2)}{40(-6 + 3 K +
              8 \pi G r_i^2\tilde{\rho}_0)},~\ldots
\end{eqnarray}

\begin{eqnarray}
d_1&=&\frac{3k(2-K)}{8\pi},
\\\nonumber
d_2 &=& \frac{k}{192\pi}[-12 K^2 + 6 (-3 + 2 (4 \pi -
k)Gr_i^2\tilde{\rho}_0) +
        K (33 + 2(-16 \pi + 3 k)Gr_i^2 \tilde{\rho}_0)],~\ldots
\end{eqnarray}
To get the value of the coefficient of $a_0$, we also use the fact
that the limit of TeVeS is GR when $k\rightarrow 0$ and
$K\rightarrow 0$. Comparing the expansion with the internal metric
of a static, spherically symmetry star in GR we have
$a_0=\frac{5}{2}-\frac{9Gm_g}{2R}-\frac{3}{2}\sqrt{1-\frac{2Gm_g}{R}}$.
At $r=R$, using Eq.(\ref{rr}), Eq.(\ref{bs}) and Eq.(\ref{lam}), we
have the relation between $m_g$ and $r_i$
\begin{eqnarray}
&&\left(K+\frac{8}{3}\pi
G\tilde{\rho}_0r_i^2\right)\frac{R^3}{r_i^2}
\\\nonumber
&-&\left\{\frac{45K^3k}{80\pi(-2 + K)} +
 \frac{18K^2(4\pi - 9k)}{80\pi(-2 + K)} -
 \frac{4K[-27k + 32 \pi^2G\tilde{\rho}_0r_i^2 + 8\pi(2 +
 3kG\tilde{\rho}_0r_i^2)]}{80\pi(-2 + K)}+
 \frac{24k(3 + 8 \pi G\tilde{\rho}_0 r_i^2)}{80\pi(-2 + K)}\right\}\frac{R^5}{r_i^4}
\\\nonumber
&+&\mathcal{O}\big(\frac{R^7}{r_i^6}\big)=2Gm_g.
\end{eqnarray}

Using the relation $\tilde g_{tt}=-e^{2\phi+\nu}$, $\tilde
g_{rr}=e^{-2\phi+\lambda}$ and $\tilde g_{\theta\theta}=\tilde
g_{\varphi\varphi}/\sin^2{\theta}=r^2e^{-2\phi}$, we have
\begin{eqnarray}
\tilde{g}_{tt}&=&-\left(\frac{5}{2}-\frac{9Gm_g}{2R}-\frac{3}{2}\sqrt{1-\frac{2Gm_g}{R}}\right)
\\\nonumber
&-&
\left(\frac{5}{2}-\frac{9Gm_g}{2R}-\frac{3}{2}\sqrt{1-\frac{2Gm_g}{R}}\right)\left[1
- \frac{3k(-2 + K)}{8\pi}\right]\left(\frac{r}{r_i}\right)^2 +
  O\left[\left(\frac{r}{r_i}\right)^4\right],
\\\nonumber
\tilde{g}_{rr}&=&1 + \left[
        K + \frac{3k(-2 + K)}{8\pi} + \frac{8}{3}\pi G\tilde{\rho}_0r_i^2\right]\left(\frac{r}{r_i}\right)^2 +
  O\left[\left(\frac{r}{r_i}\right)^4\right],
\\\nonumber
\tilde{g}_{\theta\theta}&=&r^2\
      \left\{
      \left[1+\frac{3k(-2 + K)}{8\pi}\right] \left(\frac{r}{r_i}\right)^2+O\left[\left(\frac{r}{r_i}\right)^4\right]
      \right\},
\\\nonumber
\tilde{g}_{\varphi\varphi}&=&\sin^2\theta\tilde{g}_{\theta\theta}.
\end{eqnarray}
It is clearly that the internal physical metric of spherical star in
TeVeS will get back that in GR when $k\rightarrow 0$ and
$K\rightarrow 0$.

Though the stellar model discussed in this section is not a perfect
model, it is still a first-order approximation of small star of
which the pressure is not large. The procedure above shows us that
theoretically we can deal with a real star in TeVeS in the similar
way, but the numerical calculation will be more complex.

\vspace{0.4cm} \noindent\textbf{5. Features of Motion Around a
Spherically Symmetric Object in TeVeS} \vspace{0.4cm}

Based on the analyses of the sections above, the features of motion
around any spherically symmetric compacted objects not only the
black holes can be discussed now. An analytic
solution\cite{Giannios} has been gotten in spherically symmetric
spacetime in TeVeS for arbitrary $k$ and $K$, where the metric in
TeVeS is consistent with these in general relativity, if $k=0$ and
$K=0$. The solution is based on the line element
\begin{equation}
ds^2=g_{\alpha\beta}\, dx^\alpha dx^\beta=-e^{\nu(\rho)}
dt^2+e^{\varsigma(\rho)} [dr^2+ \rho^2(d\theta^2+\sin^2\theta
d\varphi^2)], \label{ds2}
\end{equation}
which is equivalent to the line element in Eq.(\ref{ds1}).  Between
Eq.(\ref{ds1}) and Eq.(\ref{ds2}), it is easy to find the relation:
\begin{eqnarray}
e^{\nu(\rho)}&=&e^{\nu(r)},
\\\nonumber
e^{\varsigma(\rho)}d\rho^2&=&e^{\lambda(r)}dr^2,
\\\nonumber
e^{\varsigma(\rho)}\rho^2&=&r^2.
\end{eqnarray}
And the static, spherically symmetric analytic solutions are:
\begin{equation}
\phi(\rho)=\phi_c+\frac{kGm_s}{8\pi
\rho_c}\ln\left(\frac{\rho-\rho_c}{\rho+\rho_c}\right), \label{phi}
\end{equation}
\begin{equation}
e^{\nu}=\left(\frac{\rho-\rho_c}{\rho+\rho_c}\right)^{\rho_g/2\rho_c},
\label{enu}
\end{equation}
\begin{equation}
e^{\varsigma}=\frac{\big(\rho^2-\rho_c^2\big)^2}{\rho^4}\left(\frac{\rho-\rho_c}{\rho+\rho_c}\right)^{-\rho_g/2\rho_c},
\label{etheta}
\end{equation}
where
$\rho_c=\frac{\rho_g}{4}\sqrt{1+\frac{k}{\pi}\big(\frac{Gm_s}{\rho_g}\big)^2-\frac{K}{2}}$£¬
$\rho_g$ is a length scale to be determined. And $m_s$ is the
"scalar" mass which is defined by
$m_s\equiv4\pi\int_0^R(\tilde{\rho}+3\tilde{p})e^{\nu/2+3\varsigma/2-2\phi}\rho^2d\rho$,
where $R$ is radius of the matter's boundary\cite{B1}. Using the
results of post-Newtonian corrections, one has the relation between
$m_s$,$m_g$ and $\rho_g$
\begin{equation}
\rho_g-\frac{3K\rho_g^2}{8R}-\frac{kG^2m_s^2}{4\pi
R}+O(\rho_g^3/R^2)=2Gm_g,
\end{equation}
where $m_g$ is the "gravitational mass". For the sun $\rho_g/R\sim
Gm_s/R\sim10^{-5}$, one will find that $\rho_g\approx2Gm_g$ with
fractional accuracy much better than $10^{-5}$. The physical metric
may reduce to $\tilde g_{tt}=g_{tt}e^{2\phi}$ and $\tilde
g_{ii}=g_{ii}e^{-2\phi}$. And one will arrive at
\begin{equation}
\tilde g_{tt}=-\left(\frac{\rho-\rho_c}{\rho+\rho_c}\right)^a
\label{tildegtt}
\end{equation}
\begin{equation}
\tilde g_{\rho\rho}=\frac{\tilde
g_{\theta\theta}}{\rho^2}=\frac{\tilde
g_{\varphi\varphi}}{\rho^2\sin^2\theta}=\frac{\left(\rho^2-\rho_c^2\right)^2}{\rho^4}\left(\frac{\rho-\rho_c}{\rho+\rho_c}\right)^{-a}£¬
\label{tildegrr}
\end{equation}
where $a\equiv\frac{\rho_g}{2\rho_c}+\frac{kGm_s}{4\pi \rho_c}$.
When $k=0$ and $K=0$, the metric will reduce to the metric in GR.

The relation between energy and momentum for a test particle of rest
mass $m$ in curved space is
\begin{equation}
\tilde g^{\alpha\beta}p_\alpha p_\beta+m^2=0, \label{geoeq}
\end{equation}
where $p_\alpha=\partial S/\partial x^\alpha$ and $S$ is the
Hamilton-Jacobi function. Thus Hamilton-Jacobi equation for
propagation of wave crests in TeVeS spherically symmetric geometry
becomes
\begin{equation}
\tilde g^{tt}\big(\frac{\partial S}{\partial t}\big)^2+\tilde
g^{\rho\rho}\big(\frac{\partial S}{\partial\hat{\rho}}\big)^2+\tilde
g^{\rho\rho}\rho^{-2}\big(\frac{\partial
S}{\partial\theta}\big)^2+\tilde
g^{\rho\rho}\rho^{-2}\sin^{-2}\theta\big(\frac{\partial
S}{\partial\varphi}\big)^2-m^2=0, \label{HJ}
\end{equation}
where $\hat{\rho}=\sqrt{g_{rr}} r$. Set
\begin{equation}
S=-\tilde Et+S_1(\hat{\rho})+S_2(\theta)+S_3(\varphi).
\label{HJfunction}
\end{equation}
Substitute Eq.(\ref{HJfunction}) into Eq.(\ref{HJ}), we have
\begin{equation}
\tilde g^{tt}\tilde E^2+\tilde g^{\rho\rho}\big(\frac{\partial
S_1}{\partial\hat{\rho}}\big)^2+\tilde
g^{\rho\rho}\rho^{-2}\big(\frac{\partial
S_2}{\partial\theta}\big)^2+\tilde
g^{\rho\rho}\rho^{-2}\sin^{-2}\theta\big(\frac{\partial
S_3}{\partial\varphi}\big)^2+m^2=0. \label{HJ1}
\end{equation}
Solving the equation we find the solution of Hamilton-Jacobi
function
\begin{equation}
S=-\tilde Et+\int^\rho \sqrt{-\frac{\tilde g^{tt}}{\tilde
g^{\rho\rho}}\tilde E^2-\frac{m^2}{\tilde g^{\rho\rho}}-\frac{\tilde
L^2}{\rho^2}}d\hat{\rho}+\int^\theta\sqrt{\tilde
L^2-\frac{p_\varphi}{\sin^2\theta}} d\theta+p_\varphi \varphi
,\label{HJfunction1}
\end{equation}
where $\tilde L$ is the angular momentum and $\tilde E$ is the
energy of the system \cite{Li}.

Find the relation between $t$ and $\hat{\rho}$ by considering
"interference of wave crests" belonging to slight different $\tilde
E$ values
\begin{equation}
\frac{\partial S}{\partial \tilde E}=-t+\int^\rho
\frac{-\frac{\tilde g^{tt}}{\tilde g^{\rho\rho}}\tilde
E}{\sqrt{-\frac{\tilde g^{tt}}{\tilde g^{\rho\rho}}\tilde
E^2-\frac{m^2}{\tilde g^{\rho\rho}}-\frac{\tilde L^2}{\rho^2}}}
d\hat{\rho}=0. \label{trho}
\end{equation}
Then one has
\begin{equation}
\frac{d\hat{\rho}}{dt}=\frac{\sqrt{-\frac{\tilde g^{tt}}{\tilde
g^{\rho\rho}}\tilde E^2-\frac{m^2}{\tilde g^{\rho\rho}}-\frac{\tilde
L^2}{\rho^2}}}{-\frac{\tilde g^{tt}}{\tilde g^{\rho\rho}}\tilde
E}.\label{drhodt}
\end{equation}
When $d\hat{\rho} /dt =0$, Eq.~(\ref{drhodt}) infers the effective
potential energy of the system:
\begin{equation}
\tilde U=\left[-\frac{\tilde g^{\rho\rho}}{\tilde
g^{tt}}\left(\frac{m^2}{\tilde g^{\rho\rho}}+\frac{\tilde
L^2}{\rho^2}\right)\right]^{1/2} . \label{effectivepotential}
\end{equation}
Substituting the physical metric $\tilde g^{tt}$ and $\tilde
g^{\rho\rho}$ into Eq.(\ref{effectivepotential}), we have the
effective potential for unit mass
\begin{equation}
U=\left(\frac{\rho-\rho_c}{\rho+\rho_c}\right)^{a/2}
\left[1+\frac{\tilde
L^2}{m^2}\frac{\rho^2}{\left(\rho^2-\rho_c^2\right)^2}\left(\frac{\rho-\rho_c}{\rho+\rho_c}\right)^a\right]^{1/2},
\label{effectivepotential0}
\end{equation}
where $U=\tilde U/m$ is the effective potential for unit mass of the
test particle. Defining three dimensionless quantities
$\mathfrak{r}\equiv {\rho}/{\rho_g}$, $\mathfrak{r}_c\equiv
{\rho_c}/{\rho_g}$ and $L\equiv{\tilde L^2}/{(m^2\rho_g^2)}$, we get
\begin{equation}
U=\left(\frac{\mathfrak{r}-\mathfrak{r}_c}{\mathfrak{r}+\mathfrak{r}_c}\right)^{a/2}
\left[1+L^2\frac{\mathfrak{r}^2}{\left(\mathfrak{r}^2-\mathfrak{r}_c\right)^2}\left(\frac{\mathfrak{r}-\mathfrak{r}_c}{\mathfrak{r}+\mathfrak{r}_c}\right)^a\right]^{1/2},
\label{effectivepotential1}
\end{equation}
where
$a=\frac{1}{2\mathfrak{r}_c}\left(1+\frac{k}{2\pi}\frac{Gm_s}{\rho_g}\right)$
and
$\mathfrak{r}_c=\frac{1}{4}\sqrt{1+\frac{k}{\pi}\big(\frac{Gm_s}{\rho_g}\big)^2-\frac{K}{2}}$.

Energy, in units of the rest mass $m$ of the particle, is denoted
$E=\tilde E/m$. It is worth noticing that one can define a turning
point by the condition $U^2=E^2$. Obviously, the minimum and maximum
of the potential depend on the $L$. The roots of $\partial
U/\partial\rho=0$ are given in terms of the $L$ by numerical
calculation, so that we can define the lowest reduced angular
momentum $L_{crit}$ by the condition
$U_{min}(L_{lowest})=U_{max}(L_{lowest})$. In other words, there is
no periastron for $L\leq L_{lowest}$, and any incoming particle is
necessarily pulled into $\rho=\rho_c$. The radius of stable circular
orbits represents the point sitting at minimum of effective
potential. The last stable circular orbit corresponds to
$L_{lowest}$. It is easily found that $L_{lowest}$ and the radius of
stable circular orbits depend on $k$ and $K$. We also define
$L_{crit}$ by the condition $U_{max}(L_{crit})=1$, then there are
bound orbits for $L> L_{crit}$ and test particles coming in from
$\rho=\infty$ with $E^2<U_{max}^2$ reach periastrons and then return
to $\rho=\infty$, but particles from $r=\infty$ with $E^2>U_{max}^2$
get pulled into $\rho=\rho_c$. There are unstable circular orbits at
the maximum of the effective potential, in which maximum moves
outward form the radius of stable circular orbits for $L=\infty$ to
the radius of stable circular orbits for $L_{lowest}$.

Giannios\cite{Giannios} had proved that the solution for the metric
of static, spherically symmetric black holes is identical to the
Schwarzshild solution in GR. To show the difference of features of
motion around a spherically symmetric object between in TeVeS and in
GR, we can discuss the corrections of stars, such as sun, but not
black holes in TeVeS. For sun $Gm_s/\rho_g\sim10^{-10}$\cite{B1}, so
it is reasonable to neglect the terms with $Gm_s/\rho_g$ in the
expression of $a$ and $\mathfrak{r}_c$, i.e. $k$ is insensitive to
$L_{lowest}$ and $L_{crit}$. We define the relative errors of
$L_{lowest}$ and $L_{crit}$
\begin{equation}
\Delta
L_{lowest}=\frac{L_{lowest}(K)-L_{lowest}(K=0)}{L_{lowest}(K=0)},
\end{equation}
\begin{equation}
\Delta L_{crit}=\frac{L_{crit}(K)-L_{crit}(K=0)}{L_{crit}(K=0)}.
\end{equation}
With the numerical calculation, the values of $L_{lowest}$, $\Delta
L_{lowest}$ and $L_{crit}$, $\Delta L_{crit}$ for different $K$ are
listed in Table~\ref{KkLL}. Note that $K=0$ corresponds to GR. We
find $L_{lowest}$ and $L_{crit}$ have a small difference between GR
and TeVeS.

\begin{table*}
\caption{\label{KkLL} Values of $L_{lowest}$, $\Delta L_{lowest}$
and $L_{crit}$, $\Delta L_{crit}$ for different $K$}
\begin{tabular}{crrrr}
\hline\hline \vspace{0.1cm}  $K$ & $L_{lowest}$ &  $\Delta L_{lowest}$ & $L_{crit}$ & $\Delta L_{crit}$ \\
\hline
0      & 1.73205 &    0            &  2         &   0                    \\
0.01   & 1.73221 &    0.009013\%    &  2.0002    &  0.00992684\%           \\
0.02   & 1.73236 &    0.018022\%    &  2.0004    &  0.019847\%             \\
0.03   & 1.73252 &    0.027025\%    &  2.0006    &  0.0297605\%            \\
0.04   & 1.73267 &    0.036023\%    &  2.00079   &  0.0396673\%            \\
0.05   & 1.73283 &    0.045016\%    &  2.00099   &  0.0495674\%            \\
0.06   & 1.73299 &    0.054003\%    &  2.00119   &  0.059461\%             \\
0.07   & 1.73314 &    0.062985\%    &  2.00139   &  0.0693478\%            \\
0.08   & 1.73330 &    0.071961\%    &  2.00158   &  0.0792281\%            \\
0.09   & 1.73345 &    0.080932\%    &  2.00178   &  0.0891018\%            \\

0.1    & 1.73361 &    0.089899\%    &  2.00198   &  0.0989688\%            \\
0.2    & 1.73516 &    0.179271\%    &  2.00395   &  0.19728\%            \\
0.3    & 1.73669 &    0.268126\%    &  2.0059    &  0.292944\%            \\
0.4    & 1.73822 &    0.356470\%    &  2.00784   &  0.391974\%            \\
0.5    & 1.73975 &    0.444313\%    &  2.00977   &  0.488381\%            \\
0.6    & 1.74126 &    0.531661\%    &  2.01168   &  0.584175\%            \\
0.7    & 1.74276 &    0.618522\%    &  2.01359   &  0.679367\%            \\
0.8    & 1.74426 &    0.704904\%    &  2.01548   &  0.773968\%            \\
0.9    & 1.74575 &    0.790813\%    &  2.01736   &  0.867986\%            \\

1.0    & 1.74723 &    0.876256\%    &  2.01923   &  0.961433\%            \\
1.1    & 1.74870 &    0.961241\%    &  2.02109   &  1.05432\%            \\
1.2    & 1.75016 &    1.045774\%    &  2.02293   &  1.14665\%            \\
1.3    & 1.75162 &    1.129861\%    &  2.02477   &  1.23843\%            \\
1.4    & 1.75307 &    1.213508\%    &  2.02659   &  1.32968\%            \\
1.5    & 1.75451 &    1.296722\%    &  2.02841   &  1.4204\%            \\
1.6    & 1.75594 &    1.379509\%    &  2.03021   &  1.5106\%            \\
1.7    & 1.75737 &    1.461874\%    &  2.03201   &  1.6003\%            \\
1.8    & 1.75879 &    1.543824\%    &  2.03379   &  1.68948\%            \\
1.9    & 1.76020 &    1.625363\%    &  2.03556   &  1.77817\%            \\
\hline
\end{tabular}
\end{table*}

\vspace{0.4cm} \noindent\textbf{6. Conclusion} \vspace{0.4cm}

Based on relativistic MOND theory, we have studied the spacetime in
the external and internal of a semi-realistic star with a constant
density inside the star in this paper. We show the
Oppenheimer-Volkoff equation in TeVeS, and calculate the metric
coefficient and pressure inside the star to post-Newtonian order.
Therefore, one can determine the relation between $m_s$, $m_g$ and
$R$ by integrating the metric and scalar equation inside the star.
The more complex equation of state to approach the stellar structure
is left for future work. Furthermore, we study the features of a
test particle around the star in TeVeS and get the effective
potential which can determine the features of particle's motion. We
define two relative errors to discuss the corrections of TeVeS and
show that there is a small correction of TeVeS which can be tested
by the future experiments.

\section*{Acknowledgement}

This work is supported by National Science Foundation of China under
Grant No. 10473007 and No. 10503002.

\end{document}